\documentclass[twocolumn,pra,article,showpacs]{revtex4-1}
\usepackage{bm}
\usepackage{epsfig}
\usepackage{graphicx}
\usepackage{amssymb}
\usepackage{amsthm}
\usepackage{amsmath}
\usepackage{color}
\newcommand{\F}{\hat{\mathbf{F}}}
\newcommand{\Fx}{\hat{F}_x}
\newcommand{\Fy}{\hat{F}_y}
\newcommand{\Fz}{\hat{F}_z}
\newcommand{\I}{\hat{\textrm{I}}}
\newcommand{\HH}{\hat{H}}
\newcommand{\f}{\mathbf{f}}
\newcommand{\U}{\hat{U}}
\newcommand{\X}{\hat{X}}
\newcommand{\Y}{\hat{Y}}
\newcommand{\beq}{\begin{equation}}
\newcommand{\enq}{\end{equation}}

\begin{document}

\title{Stability of nonstationary states of spin-$1$ Bose-Einstein condensates}

\author{H. M\"akel\"a${}^1$, M. Johansson${}^2$, M. Zelan${}^{1,3}$, and E. Lundh${}^1$}
\affiliation{${}^1$Department of Physics, Ume\aa \,\,University, SE-901 87 Ume\aa, Sweden}
\affiliation{${}^2$Department of Physics, Chemistry and Biology, Link\"oping \,\,University, SE-581 83 Link\"oping, Sweden}
\affiliation{${}^3$Joint Quantum Institute, National Institute of Standards and Technology and University of Maryland, 
 Gaithersburg, MD, 20899, USA}
\date{September 8, 2011}

\begin{abstract}
The stability of nonstationary states of homogeneous spin-1 Bose-Einstein condensates is studied  by performing Bogoliubov analysis in a frame of reference where the state 
is stationary. In particular, the effect of an external magnetic field  
is examined. It is found that a nonzero magnetic field introduces instability 
in a $^{23}$Na condensate. The wavelengths of this instability can be controlled 
by tuning the strength of the magnetic field.  
In a $^{87}$Rb condensate this instability is present already at zero magnetic field.  
Furthermore, an analytical bound for the size of a stable condensate is found,  and 
a condition for the validity of the single-mode approximation is presented.   
Realization of the system in a toroidal trap is discussed and the full time 
development is simulated. 
\end{abstract}


\maketitle

\section{Introduction}\label{Sec:Introduction}
The excitations and stability of spinor Bose-Einstein condensates (BECs) have been 
the subject of intense study in recent years. The topic was first explored 
 in two seminal theoretical papers in 1998 \cite{Ho98,Ohmi98}. These papers discussed the stability of $F=1$ BECs against small external perturbations, such as fluctuations in the trapping potential or small magnetic field gradients. In an unstable spinor condensate small perturbations may lead to domain formation, where the populations of spin components become position dependent. 
The theoretical studies typically examine the stability of stationary states. This is done using a linear stability analysis, where a small perturbation is added  
to the stationary state and the time evolution equations are expanded to first order in the perturbation \cite{Ho98,Ohmi98,Ueda00,Robins01,Ueda02,Murata07}. These theoretical studies have shown that the stationary states of ferromagnetic $F=1$ condensates may be unstable. 
Antiferromagnetic condensates in stationary states, on the other hand, appear to be stable against small perturbations in the absence of external fields.

The properties of spinor condensates have been studied by various experimental groups (see, for example, \cite{Stenger98,Chang04,Chang05,Kronjager05,Sadler06,Black07,Leslie09,Kronjager10,Guzman11}). 
Signs of instability have been observed by the Chapman group \cite{Chang05} in 
a ferromagnetic $F=1$ condensate and the Sengstock group in an antiferromagnetic phase of an $F=2$ rubidium condensate \cite{Kronjager10}. 
The group of Stamper-Kurn has observed that an $F=1$ $^{87}$Rb condensate prepared in a stationary paramagnetic state develops spin textures as it is rapidly quenched across 
a quantum phase transition \cite{Sadler06,Leslie09}. They saw a similar phenomenon when an unmagnetized rubidium gas was cooled to quantum degeneracy \cite{Guzman11}. 
Experiments often concentrate on spin-mixing  dynamics, 
which can be initiated by preparing the condensate in a non-stationary state.  
There are a few theoretical studies where the stability analysis has been extended to nonstationary states: The effects of a non-zero magnetic field on the stability of states with time-independent spin populations but oscillating relative phases have been examined in Refs.\ \cite{Matuszewski08,Matuszewski09,Matuszewski10}.  
Although the spin populations remain constant throughout the time evolution, these states are nonstationary because the relative phases of the spin components vary in time. Another example is given by Ref.\ \cite{Zhang05}, where the stability of states that show oscillations both  in spin populations and relative phases was studied, under the assumption that the magnetic field vanishes.

In the present paper, we generalize these findings to arbitrary states with and without magnetic fields.  
We concentrate on the case where the magnetic field is nonzero and also consider a situation where the spin populations, as well as the phases of the spin components, oscillate in time. In the analytical calculations we assume that the particle density is homogeneous.  
We solve analytically for the eigenmodes and eigenstates of a certain class of nonstationary states in rubidium and sodium condensates in an arbitrary magnetic field. 
We show that this can be done in a simple way by transforming to a reference frame where the states in question are stationary. In this way, the time dependence of the matrix determining the stability properties can be eliminated, and the problem becomes easily tractable. 
We emphasize that we are studying the stability in nonzero magnetic field. The zero-field case has been discussed in \cite{Zhang05}. 
Knowing the eigenmodes allows 
us to derive an analytical formula that connects the stability of a condensate to its size and 
the strength of the magnetic field. 
In \cite{Matuszewski08} it was found that a sodium condensate is unstable at a low magnetic field provided that the condensate is larger than the spin healing length. 
We show here that in a situation where the magnetic-field energy dominates over the spin interaction energy, both rubidium and sodium condensates may be unstable even when the size of the condensate is smaller than the spin healing length.

This paper is organized as follows. Section \ref{sec:overview} introduces the system and presents the 
Hamiltonian and time evolution equations. Sec. \ref{sec:stability} formulates the theory of the Bogoliubov analysis of nonstationary states. In Sec. \ref{sec:analytical} 
analytical results concerning the 
stability of $F=1$ condensates are derived. In Section \ref{sec:orthogonal} the stability of a state orthogonal to the magnetic field is studied using the results of Floquet theory. Additionally, a sufficient condition for the size of a stable condensate is derived and a condition for the validity of the single-mode approximation is presented.  
In Sec. \ref{sec:ring} the realization of instabilities in a condensate confined in a toroidal trap is discussed. The time development is simulated using the Gross-Pitaevskii equation. Finally, Sec. \ref{sec:conclusions} contains the concluding remarks.

\section{Theory of a spin-1 condensate}\label{sec:overview}

The order parameter of a spin-$1$ Bose-Einstein condensate can be written as $\psi=(\psi_{1},\psi_{0},\psi_{-1})^T$, where $T$ denotes transpose. The normalization is now chosen as $\sum_{m=-1}^1|\psi_{m}|^2=n$, where $n$ is the total particle density. 
We assume that the trap confining the condensate is such that all the
components of the hyperfine spin can be trapped simultaneously and are degenerate 
in the absence of magnetic field.  
If the system is exposed to an external magnetic field that  
is parallel to the $z$ axis, the energy functional reads 
\begin{align}
\nonumber
E[\psi] =\!\! \int d{\bm r} 
\big\{ &\psi^\dag(\mathbf{r}) \hat{h}\psi(\mathbf{r})  
 +\frac{1}{2}\{g_0 [\psi^\dag(\mathbf{r})\psi(\mathbf{r})]^2  + g_2 \langle\F\rangle^2\}\\
 &- p\langle \Fz\rangle + q\langle \Fz^2\rangle\big\},
\label{energy_functional(f=1)}
\end{align}
where $\hat{h}= -\frac{\hbar^2 \nabla^2}{2m} + U(\mathbf{r})  -\mu$ , 
$\F=(\Fx,\Fy,\Fz)$ is the spin operator of a spin-1 particle and
we use the notation $\langle  \mathbf{X} \rangle=\psi^\dag(\mathbf{r}) \mathbf{X}\psi(\mathbf{r})$.  
Here $U$ is the external trapping potential and the chemical potential, taking care of the conservation of the total particle number, is denoted by $\mu$.   
The strength of the spin-independent interaction is characterized by $g_0=4\pi \hbar^2(a_0+2a_2)/3m$, while $g_2=4\pi \hbar^2(a_2-a_0)/3m$ describes the spin-dependent scattering.  Here $a_F$ is the $s$-wave scattering length for two atoms colliding with total angular momentum $F$. 
For ${}^{87}$Rb the scattering lengths used in this paper are $a_0=101.8a_B$ and $a_2=100.4 a_B$ \cite{vanKempen02} with $a_B$ being the Bohr radius. For ${}^{23}$Na the corresponding values are $a_0=50.0a_B$ and $a_2=55.1a_B$ \cite{Crubellier99}. 
(Note, however, that there are many estimates for the difference $a_2-a_0$ in the literature \cite{Stenger98,Black07,Crubellier99,Burke98}.) 
The magnetic field introduces two terms, one of which is given by the 
linear Zeeman term  $p=-g\mu_{\rm B}B$, where $g$ is the Land\'e hyperfine $g$ factor, 
 $\mu_{\rm B}=e\hbar/2m_{\rm e}$ is the Bohr magneton ($m_{\rm e}$ is the electron mass, 
and $e>0$ is the elementary charge), and $B$ is the external magnetic field. The other term is  the quadratic Zeeman term 
\begin{eqnarray}
q=\frac{(g\mu_{\rm B}B)^2}{E_{\rm hf}},
\end{eqnarray}
where $E_{\rm hf}$ is the hyperfine splitting. 
For $^{87}$Rb and $^{23}$Na the hyperfine splittings are $E_{\rm hf}= 6.835$~GHz and $E_{\rm hf}= 1.772$~GHz, respectively. In both cases $g=-1/2$. The value of $q$ can be made negative by using a linearly polarized microwave field \cite{Gerbier06}. In this paper we concentrate on non-negative $q$.

We characterize the spin of the state $\psi$ by the spin vector $\f$, defined as 
\begin{align}
\f(\mathbf{r})=  \frac{\psi^\dag(\mathbf{r}) \F \psi(\mathbf{r})}{n(\mathbf{r})}.
\end{align}
The length of this vector is denoted by $f$, $f=||\f||$.
In addition to the number of particles, the magnetization in the $z$ direction, defined as 
\begin{align}
\label{Mz}
M_z= \frac{\int d\mathbf{r}\,n(\mathbf{r}) f_z (\mathbf{r})}{\int d\mathbf{r}\,n(\mathbf{r})},
\end{align}
is also a conserved quantity. The Lagrange multiplier related to magnetization  
can be included into $p$. We consider mostly homogeneous systems, for which $M_z=f_z$.   
The time evolution is governed by 
\begin{align}
\label{time-evolution}
i\hbar\frac{\partial}{\partial t}\psi(t)=\HH [\psi(t)]\psi(t), 
\end{align}
where the Hamiltonian is defined  as 
\begin{align}
\label{Hamiltonian}
\HH [\psi]=[\hat{h}+g_0 n(\mathbf{r})]\I  +g_2 \langle\F\rangle\cdot \F -p\Fz + q\Fz^2. 
\end{align}
For a homogeneous system $\hat{h}\rightarrow -\mu$, and the density $n$ becomes position independent. Consequently the energy of a homogeneous system reads 
\begin{align}
\label{Ehomo}
E[\psi] &=-\mu+\frac{1}{2}(g_0 n+ g_2 n f^2)-p f_z + q/n\langle \Fz^2\rangle.
\end{align}

In the following analysis the time evolution operator $\U_\psi$ of the state $\psi$, $\psi(t)=\U_\psi(t)\psi(0)$, is used frequently. 
This operator can be formally written as 
\begin{align}
\label{Ut}
\U_\psi(t)=\hat{T} e^{-i/\hbar \int_{0}^t d\tau\, \HH [\psi(\tau)]},
\end{align}
where $\hat{T}$ is a time-ordering operator. Note that the Hamiltonian appearing in the 
exponent depends on the state of the system. In some cases $\U_\psi$ can be solved analytically, but in general, numerical calculations are necessary. 
In this paper  the numerical calculation is done by first solving the time evolution of $\psi$, with the help of which we get the time-dependent Hamiltonian. The columns of the propagator $\U_\psi$ can then be obtained by calculating the time evolution (under $\HH$) 
of the basis states $(1,0,0)^T$, $(0,1,0)^T$ , and $(0,0,1)^T$.

\section{Stability of nonstationary states}\label{sec:stability}

We study the stability of nonstationary states using Bogoliubov analysis. 
This is done in a basis where the state we are interested in is time-independent.  
We define a new (time-dependent) basis  $\{|+1\rangle^{\textrm{new}},|0\rangle^{\textrm{new}}, |-1\rangle^{\textrm{new}}\}$ in terms of the old basis $\{|+1\rangle,|0\rangle, |-1\rangle\}$ as
 $|\nu\rangle^{\textrm{new}}=\U_\psi^{-1}|\nu\rangle$, $\nu=+1,0,-1$. 
Here $\U_\psi$ is defined as in Eq. (\ref{Ut}).  
 In the new basis, the energy of an arbitrary  
state $\phi$ is given by 
\begin{align}
\label{Erot}
E^{\textrm{new}}[\phi]&=E[\U_\psi\phi]+i\hbar\langle\phi|\left(\frac{\partial}{\partial t}\U^{-1}_\psi\right)\U_\psi\phi\rangle, 
\end{align}
and the time evolution of $\phi$ can be obtained from the equation  
\begin{align}\label{variE}
i\hbar\frac{\partial\phi}{\partial t}=\frac{\delta E^{\textrm{new}}[\phi]}{\delta\phi^\dag}. 
\end{align}
Equation (\ref{Erot}) can be simplified using the equation  
$i\hbar \left(\frac{\partial}{\partial t} \U_\psi^{-1}\right) \U_\psi = -\U_\psi^{-1} \HH [\psi] \U_\psi$. Using Eqs. (\ref{Ehomo}), (\ref{Erot}) and (\ref{variE}) it is then easy to see that the state $\phi=\psi(0)$ does not evolve in time, confirming that $\psi(0)$ is a stationary state in the new frame.   
We study the stability of $\psi(0)$ by replacing $\psi(0)\rightarrow \psi(0) +\delta\psi$ in the time evolution equations obtained from Eq. (\ref{variE}) and expand the resulting 
equations to first order in $\delta\psi$. The perturbation $\delta\psi=(\delta\psi_1,\delta\psi_0,\delta\psi_{-1})^T$ is assumed to be of the form 
\begin{align*}
\delta\psi_j=\sum_{\mathbf{k}} \left[ u_{j;\mathbf{k}}(t)\,e^{i\mathbf{k}\cdot\mathbf{r}}
-v_{j;\mathbf{k}}^{*}(t)\, e^{-i\mathbf{k}\cdot\mathbf{r}}\right],\quad j=-1,0,1.
\end{align*}
Straightforward calculation gives the differential equation for the time evolution of the  perturbations as
\begin{align}
&i\hbar\frac{\partial}{\partial t}\begin{pmatrix}
u_{1;\mathbf{k}}\\
u_{0;\mathbf{k}}\\
u_{-1;\mathbf{k}}\\
v_{1;\mathbf{k}}\\
v_{0;\mathbf{k}}\\
v_{-1;\mathbf{k}}
\end{pmatrix}
=\HH_{B}
\begin{pmatrix}
u_{1;\mathbf{k}}\\
u_{0;\mathbf{k}}\\
u_{-1;\mathbf{k}}\\
v_{1;\mathbf{k}}\\
v_{0;\mathbf{k}}\\
v_{-1;\mathbf{k}}
\end{pmatrix},\\ 
\label{HBG}
&\HH_{B}=\begin{pmatrix}
\X &- \Y\\
\Y^* & -\X^*
\end{pmatrix},
\end{align}
where the $3\times 3$ matrices $\X$ and $\Y$ are defined as  
\begin{align}
\nonumber
\X =&\,\, \epsilon_k + g_0 |\psi(0)\rangle\langle\psi(0)|\\
\label{X}
&+g_2 \sum_{\nu=x,y,z} |\U_\psi^\dag(t)\hat{F}_\nu\psi(t)\rangle\langle \U_\psi^\dag(t)\hat{F}_\nu\psi(t)| \\
\nonumber
\Y =&\,\, g_0 |\psi(0)\rangle\langle\psi^*(0)| \\
\label{Y}
&+g_2 \sum_{\nu=x,y,z} |\U_\psi^\dag (t)\hat{F}_\nu\psi(t)\rangle\langle [\U_\psi^\dag (t)\hat{F}_\nu \psi(t)]^*|, \\
\label{epsilonk}
\epsilon_k \equiv & \frac{\hbar^2 k^2}{2m}, 
\end{align}
and $\psi(t)=\U_\psi(t)\psi(0)$. In the rest of the paper we call the operator $\HH_{B}$ the Bogoliubov matrix. The magnetic field dependence appears in the Bogoliubov matrix through the magnetic field dependence of $\U_\psi$. 
The operator $\HH_{B}$ is typically time-dependent and 
the time evolution of the perturbations is given by the time-ordered integral 
\begin{align}
\label{Upertt}
\U_{B}(t)=\hat{T} e^{-i/\hbar \int_{0}^t d\tau\, \HH_{B}(\tau)}.
\end{align}
In general, both $\U_\psi$ and $\U_{B}$ have to be calculated numerically. 
In some special cases it is possible to express $\U_{B}$ analytically in terms of a time-independent Bogoliubov matrix, and the stability can be determined by calculating the eigenvalues of this matrix. The system is unstable if at least one of the eigenvalues of $\hat{H}_B$ has a nonzero complex part. Another case considered in this paper is one where the time evolution of  $\hat{H}_B$ is periodic. This makes it possible to use Floquet theory to study the stability.  
We first discuss some special cases that allow analytical solution, and then proceed to the case where $\hat{H}_B$ is periodic.

\section{Analytical results}\label{sec:analytical}
In this section the stability is studied using mainly analytical means. 
First we analyze the stability of a system where the spin and magnetic field are parallel 
in the initial state. In the second case we concentrate on the stability in the limit of a large magnetic field. 

\subsection{Parallel spin and magnetic field}
One case where the stability can be studied analytically is a system 
where the spin and magnetic field are parallel in the initial state,    
 $\langle \Fx\rangle=\langle \Fy\rangle=0$. 
It is easy to show that the state has to be of the form 
\begin{align}
\label{psipara}
\psi_\parallel = \sqrt{n}\begin{pmatrix}
\sqrt{(1+f_z)/2}\\
0\\
\sqrt{(1-f_z)/2}
\end{pmatrix},\quad |f_z|= f,  
\end{align}
where the relative phase of the two nonzero spin components can be chosen to be zero 
due to the fact that the energy is invariant under rotations around the $z$ axis. 
In general, the stability properties of two states that can be obtained from each other using 
an element of the symmetry group of the energy are identical 
\footnote{This can be proven as follows. 
If $\hat{V}$ is an element of the (now unitary) symmetry group of the energy, 
then $\hat{U}_{\hat{V}\psi}=\hat{V}\hat{U}_\psi\hat{V}^{\dag}$. By replacing $\psi(0)\rightarrow \hat{V}\psi(0)$ and $\hat{U}_\psi\rightarrow \hat{V}\hat{U}_\psi\hat{V}^{\dag}$ in Eqs. (\ref{X}) and  (\ref{Y}) we find that  $\hat{X}\rightarrow \hat{V} \hat{X}\hat{V}^\dag$ and 
 $\hat{Y}\rightarrow \hat{V} \hat{Y}\hat{V}^T$. Consequently,  
 $\hat{H}_B\rightarrow \hat{W}\hat{H}_B\hat{W}^\dag$, where $\hat{W}$ is a block diagonal matrix 
 $\hat{W}=\textrm{diag}(\hat{V}\,\, \hat{V}^*)$. Because $\hat{H}_B$ and  $\hat{W}\hat{H}_B\hat{W}^\dag$ have the same eigenvalues, they also have identical stability properties.}. Therefore, instead of studying the stability of all possible states, 
it is enough to concentrate on those states that cannot be connected by an element of the  symmetry group. We remark that the stability analysis presented in this section is valid for all states at zero magnetic field. In this  case we can make use of the fact that for any spin state $\psi$ there exists a spin rotation operator 
$R(\alpha,\beta,\gamma)\equiv e^{-i\alpha \Fz}e^{-i\beta \Fy}e^{-i\gamma \Fz}$ such that 
$\psi=e^{i\tau}R(\alpha,\beta,\gamma)\psi_\parallel$, where $(\alpha,\beta,\gamma)$ are the Euler angles and $\tau$ is the global phase.  
At zero magnetic field the initial state can therefore always be assumed to be of the form $\psi_{\parallel}$ given in Eq. (\ref{psipara}).

\begin{figure}[h]
\begin{center}
\includegraphics[scale=.9]{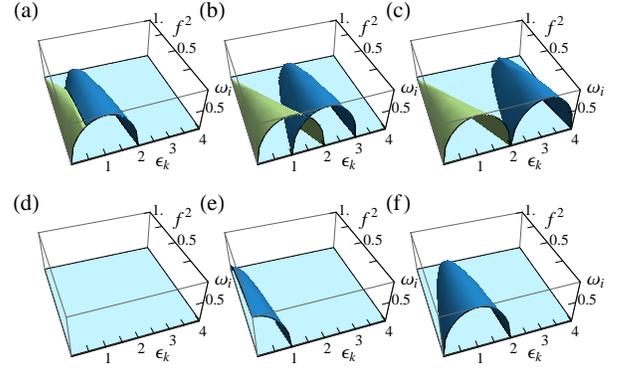} 
\end{center}
\caption{(Color online) The amplitude of the unstable frequencies $\omega_\textrm{i}=\text{Im}[\omega]$ 
in the case $\f\parallel\mathbf{B}$ for (a)-(c) rubidium and (d)-(f) sodium. The units of $\epsilon_k$ and $\omega_\textrm{i}$ are $|g_2|n$  and $|g_2|n/\hbar$, respectively.  
Here in (a) and (d) $q=0$, in (b) and (e) $q=|g_2|n$, and in (c) and (f) $q=2|g_2|n$.   
The green color (left lobes in top row) indicates the unstable modes given by Eq. (\ref{omega34}) 
[here called magnetization modes; see (\ref{psi34})], while the blue color gives the instability arising from the modes of Eq. (\ref{omega56}) [now called spin modes; see (\ref{psi56})]. 
The region corresponding to $f=0$ in the bottom row agrees with the results presented in
 Fig. 4(b) of Ref. \cite{Matuszewski10}. 
\label{fig:parallel}}
\end{figure}

In Appendix \ref{appendixA} we show that the spin populations of $\psi_\parallel(t)=\U_{\psi_\parallel}(t)\psi_\parallel$ are time independent regardless of the value of $q$.  Then Eq. (\ref{Hamiltonian}) gives the propagator   
\begin{align}
\U_{\psi_\parallel}(t) =e^{-it(g_0 n-\mu)/\hbar } e^{-i t [(g_2 n f_z-p) \Fz + q\Fz^2]/\hbar},
\end{align}
and the matrices appearing in the Bogoliubov Hamiltonian become 
\begin{align}
\nonumber
\X^{\parallel}=&\,\,\epsilon_k\, \I + g_0 |\psi_\parallel(0)\rangle\langle\psi_\parallel(0)|\\
&+g_2 |\Fz\psi_\parallel(0)\rangle\langle \Fz\psi_\parallel(0)|
+g_2 n (\I-\Fz^2)\\
\nonumber 
\Y^{\parallel} =&\,\, g_0 |\psi_\parallel(0)\rangle\langle\psi_\parallel(0)| 
+g_2 |\Fz\psi_\parallel(0)\rangle\langle \Fz\psi_\parallel(0)| \\
&+ e^{-i 2 q t /\hbar} g_2 n \sqrt{1-f_z^2} (\I-\Fz^2). 
\end{align}
The Bogoliubov matrix $\HH_{B}^\parallel$ is such that the time evolution 
of $\{u_{0;\mathbf{k}},v_{0;\mathbf{k}}\}$ is decoupled from the time evolution of $\{u_{1;\mathbf{k}},u_{-1;\mathbf{k}},v_{1;\mathbf{k}},v_{-1;\mathbf{k}}\}$. Moreover, the Bogoliubov matrix giving the time evolution of $\{u_{1;\mathbf{k}},u_{-1;\mathbf{k}},v_{1;\mathbf{k}},v_{-1;\mathbf{k}}\}$ is time-independent and 
the time-dependence of the $\{u_{0;\mathbf{k}},v_{0;\mathbf{k}}\}$ -part can be eliminated 
 by defining a new basis $\tilde{u}_{0;\mathbf{k}}=e^{i q t/\hbar}v_{0;\mathbf{k}},
\tilde{v}_{0;\mathbf{k}}=e^{-i q t/\hbar}u_{0;\mathbf{k}}$, 
and $\tilde{u}_{j;\mathbf{k}}=u_{j;\mathbf{k}},\tilde{v}_{j;\mathbf{k}}=v_{j;\mathbf{k}}$ 
 for $j=\pm 1$. After this the eigenvalues can be easily calculated 
\begin{align}
\label{omega12}
(\hbar\omega_{1,2})^2&=\epsilon_k\left[(g_0+g_2) n+\epsilon_k+n\sqrt{(g_0-g_2)^2+4 g_0 g_2 f_z^2}\right],\\
\label{omega34}
(\hbar\omega_{3,4})^2&=\epsilon_k\left[(g_0+g_2)n+\epsilon_k-n\sqrt{(g_0-g_2)^2+4 g_0 g_2 f_z^2}\right],\\
\label{omega56}
(\hbar \omega_{5,6})^2 
&=(g_2n)^2(f_z^2-1)+(\epsilon_k+g_2 n-q)^2.
\end{align} 
For $q=0$ the eigenvalues (\ref{omega12})-(\ref{omega56}) reduce to those given in \cite{Zhang05}. 
We assume that $g_0>0$ and $|g_2|\ll g_0$, which is the case both for rubidium and sodium.  
Now $\omega_{1,2}$ are always real, but $\omega_{3,4}$ can be complex if $g_2<0$;  
the unstable states lie inside a triangular region in the $(\epsilon_k,f_z^2)$ plane; 
see Figs.\ \ref{fig:parallel}(a)-\ref{fig:parallel}(c).  
For fixed values of $\omega_{5,6}$ and $q$, equation (\ref{omega56}) determines an ellipsoid 
in the $(\epsilon_k,f_z)$ plane. 
The unstable states lie in the interior of the 
ellipsoid obtained by setting $\omega_{5,6}=0$ and are constrained by the inequalities  $\epsilon_k,f^2\geq 0$; see  Figs.\ \ref{fig:parallel}(a)-(c). 
For $g_2>0$ the region of instability is shifted by $2g_2 n$ with respect to that of the $g_2<0$ system, as can be seen from Fig.  \ref{fig:parallel}. 
We see that $\psi_\parallel$ is unstable in a rubidium condensate 
if $|f_z|<1$. The same applies in a sodium condensate if $q\geq g_2n$. When $q< g_2 n$, this state is unstable if $f_z^2< -q^2+2q$. At $|f_z|=1$ the system is stabilized by the conservation of magnetization.

\begin{figure}[h]
\begin{center}
\includegraphics[scale=.95]{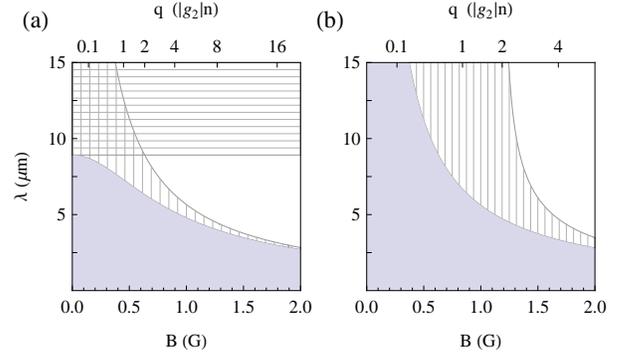} 
\end{center}
\caption{The wavelengths of the unstable perturbations in the case $\mathbf{B}\parallel \f$ for (a) rubidium and (b) sodium. Horizontal (vertical) lines denote magnetization (spin) modes. 
We have chosen $n=4\times 10^{14}$cm${}^{-3}$ and $f=0$. The latter choice gives the largest possible interval of unstable wavelengths. The shaded region gives condensate sizes, which correspond to stable 
systems regardless of the initial state; see Sec. \ref{sec:orthogonal}.
\label{fig:wavelengths}}
\end{figure}

Regardless of the sign of $g_2$, the fastest-growing unstable mode is located at $\epsilon_k=q-g_2 n$  and corresponds to the wavelength
\begin{align}
\lambda =\frac{2\pi \hbar}{\sqrt{2m(q-g_2 n)}}.
\end{align}
For a sodium condensate in a magnetic field $q<g_2n$ the fastest-growing mode is at $\epsilon_k=0$. 
In Fig. \ref{fig:wavelengths} we show the possible wavelengths of unstable perturbations 
as a function of the magnetic field. We have chosen $n=4\times 10^{14}$cm${}^{-3}$.  

The eigenvectors $\{\mathbf{x}_{j;\mathbf{k}}\}$ corresponding to the eigenvalues (\ref{omega12})-(\ref{omega56}) can be calculated analytically and are given in Appendix \ref{appendixB}. 
Using the analytical expressions for the eigenvectors the corresponding spin states can be calculated straightforwardly; see Eqs. (\ref{x1234}) and (\ref{x56}).  We denote by  $\delta\psi^i$ the state corresponding to eigenvector $i$. 
We find to lowest order in $g_2/g_0$ (see Appendix \ref{appendixB})
\begin{align}
\label{psi12}
\delta\psi^{1,2} &\approx \sum_{\mathbf{k}}  C^{1,2}(\mathbf{k}\cdot\mathbf{r},t)
\begin{pmatrix}
\sqrt{(1+f_z)/2}\\0\\\sqrt{(1-f_z)/2}
\end{pmatrix},\\
\label{psi34}
\delta\psi^{3,4} &\approx \sum_{\mathbf{k}}  C^{3,4}(\mathbf{k}\cdot\mathbf{r},t)
\begin{pmatrix}
\sqrt{(1-f_z)/2}\\0\\-\sqrt{(1+f_z)/2}
\end{pmatrix},\\
\label{psi56}
\delta\psi^{5,6} &= \sum_{\mathbf{k}}  C^{5,6}(\mathbf{k}\cdot\mathbf{r},q,t)
\begin{pmatrix}
0\\1\\0
\end{pmatrix},
\end{align} 
where $C^{j,j+1}$ contain all position, time, and magnetic field dependence. 
Of these, $\delta\psi^{1,2}$ corresponds to a change in density, while 
the magnetization, defined as in Eq. (\ref{Mz}),
 and spin direction remain unchanged. We therefore call it 
a density mode. The perturbations $\delta\psi^{3,4}$, now called magnetization modes,
 affect the density and magnetization but not the spin direction. 
Finally, $\delta\psi^{5,6}$  change the density and spin direction but not the 
magnetization and are called spin modes. 
The density modes are always stable, reflecting the fact that the spin-independent 
interaction is now repulsive. 
For $g_2<0$ the magnetization mode can be unstable, whereas for $g_2>0$ 
it is always stable. This can be understood by 
looking at how the energy behaves when the system breaks into regions 
with different spin values. Neglecting constant terms, the energy 
of an arbitrary state can be written as 
\begin{align}
E=\frac{1}{2} g_2 n f^2 +q(1-\rho_0). 
\end{align}
In the initial state $\psi_\parallel$ the energy reads
\begin{align}
E_\parallel=\frac{1}{2} g_2 n f_z^2 +q. 
\end{align}
Assume that in a region of length $L_1$ $(L_2)$ the expectation value of 
the spin in the $z$ direction is $f_{z1}$ $(f_{z2})$. 
The length of the spin vector in the $(x,y)$ plane is denoted by $f_{\perp 1}$ and 
$f_{\perp 2}$. We choose $\rho_0=0$ as the magnetization modes do not populate the zero component. Consequently, $f_{\perp 1}=f_{\perp 2}=0$, and  
taking into account the conservation of magnetization, we obtain the equations 
\begin{align}
\label{tildeE}
E &=\frac{1}{2}g_2 n\frac{L_1 f_{z1}^2+L_2 f_{z 2}^2}{L_1+L_2}+q,\\
\label{fconservation}
f_z&= \frac{L_1 f_{z1}+L_2 f_{z2}}{L_1+L_2}.
\end{align}
Without loss of generality, we choose $f_z>0$, $f_{z1}\geq f_z$, and $f_{z2}\leq f_z$. 
With the help of Eqs. (\ref{tildeE}) and (\ref{fconservation}) we obtain $E=g_2n f_z^2(x_1+x_2-x_1 x_2)/2+q$, where $x_i=f_{zi}/f_z$. Taking into account that  
$ x_1\geq 1$ and $x_2\leq 1$, we find that $x_1+x_2-x_1 x_2\geq 1$. 
Hence, for rubidium 
$E\leq E_\parallel$ and domain formation is energetically allowed. Conversely, for sodium 
 $E\geq E_\parallel$ and region formation is forbidden for energetic reasons.
Here we have neglected the contribution from the kinetic energy. 
The energy cost caused by the kinetic energy    
allows only structures with long enough wavelength compared to the 
energy gained from the interaction energy. 
When $f\approx 1$ in the initial state of a rubidium condensate, this energy gain 
is very small and allows only structures with a very long wavelength. 
This qualitative result agrees with Fig.\ \ref{fig:parallel}(a)-(c).

The spin mode (\ref{omega56}) increases the population of the zero component.  
Hence we assume domains such that $\rho_0=1$ $(f_{\perp 1}=f_{z1}=0)$ and 
$\rho_0=f_{\perp 2}=0,f_{z2}=1$. As before, we have also chosen $f_z\geq 0$. 
We get 
\begin{align}
E-E_\parallel=(1-f_z)\left(\frac{1}{2} g_2 n f_z-q \right).
\end{align}
For rubidium this is negative regardless of the value of $q$, and domain formation 
is possible. For sodium the magnetic field has to be nonzero for instability 
to appear. As $q$ increases, the energy difference $E-E_\parallel$    
grows. This excess energy is transferred into kinetic energy of the 
domain structure. For large enough $q$ this kinetic energy has a finite 
minimum value, and consequently, the wavelengths of the unstable perturbations are bounded from  above \footnote{In a sodium condensate, instead of creating domains, it may be preferable   
to increase $\rho_0$ (and thus also $f$) at a low magnetic field. This corresponds to  
a spin mode with $k=0$. This mode is not present in rubidium. 
This can be understood by noting that an increase in $\rho_0$ leads to an increase in $f$. In a rubidium condensate this decreases the spin interaction energy, which, together with decreasing magnetic field energy, leads to an energy surplus that is transferred into kinetic energy. In a sodium condensate 
increasing $f$ increases the spin interaction energy. In some cases this matches exactly the energy released from the quadratic Zeeman term, producing a mode with $k=0$.}. 
This is illustrated by Figs. \ref{fig:parallel}(b) and \ref{fig:parallel}(c). 

\subsection{Stability when $q \gg |g_2|n,\epsilon_k$}
Another case where it is possible to obtain analytical results concerning the stability of the 
system is when  $q \gg |g_2|n,\epsilon_k$. 
The relevant parameters characterizing the spin states can be determined by 
writing the general spin state as 
\begin{align}
\psi_{\textrm{gen}}=\sqrt{n}e^{i\tau}e^{-i\alpha \Fz}e^{-i\beta \Fy}e^{-i\gamma \Fz}\psi_\parallel.
\end{align} 
Here $\beta$ gives the angle between the $z$ axis and the spin direction.   
The global phase $\tau$ is irrelevant and will be set to zero. Furthermore, due to the invariance of the energy  in rotations around the $z$-axis, we can choose $\alpha=0$. The important parameters are then  $\beta$ and $\gamma$. 
In Appendix \ref{appendixA} we derive an approximate propagator for the system in the limit 
$q\gg |g_2|n$. It is given, up to a time-dependent phase, by  
\begin{align}
\label{Ulimit}
\hat{U}_{\psi_\textrm{gen}}(t)= 
e^{-i t [(g_2 n \cos\beta f -p)\hat{F}_z+(g_2 n(2\rho_0-1)+q)\hat{F}_z^2]/\hbar},
\end{align}
where $\rho_0$ is the initial population of the $|0\rangle$ component. 
When analyzing the stability as a function of $\beta$ and $\gamma$, it is important to note that 
fixing the direction of the spin does not fix the populations:
The spin direction is determined by $\beta$, while $\gamma$ 
controls the populations of the spin components. 
In more detail, 
\begin{align}
\label{rho0}
\rho_0=\frac{1}{2}[1-\sqrt{1-f^2}\cos(2\gamma)]\sin^2\beta.
\end{align}  
Now $\beta$ and $\gamma$ can be chosen to lie in the interval $[0,\pi/2]$ as     
the stability properties are identical for states corresponding to $\beta$ and $\pi-\beta$ and similarly for $\gamma$. For fixed $\beta$ and $f$, the population $\rho_0$ 
is minimized (maximized) when $\gamma=0$ ($\gamma=\pi/2$). 

\begin{figure}[h]
\begin{center}
\includegraphics[scale=.85]{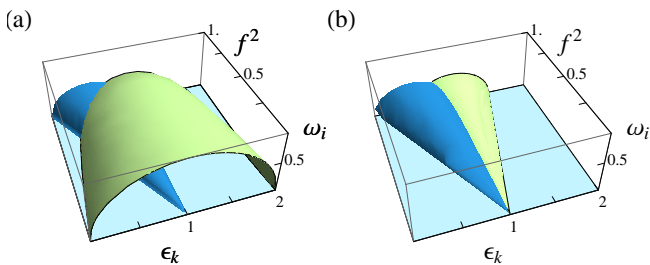} 
\end{center}
\caption{(Color online) The amplitude of long wavelength instabilities for rubidium in the limit 
$q\gg \epsilon_k,|g_2|n$. Sodium condensate does not have long wavelength instabilities in this limit. The units of $\epsilon_k$ and 
 $\omega_\textrm{i}$ are $|g_2|n$  and $|g_2|n/\hbar$, respectively. Now $\f\perp\mathbf{B}$
and the green color [larger lobe in (a) and rightmost lobe in (b)] indicates magnetization  modes, 
while the blue color indicates spin modes.  
In (a) $\gamma=0$ and in (b) $\gamma=\pi/2$. At $f=1$ 
the figures are identical.    
\label{fig:limit}}
\end{figure}

Using Eqs. (\ref{X}),(\ref{Y}), and (\ref{Ulimit}) we obtain a Bogoliubov matrix where 
the time dependence appears via terms of the form $e^{\pm 2iqt/\hbar}$. 
We use the rotating wave approximation and set these terms equal to zero. This approximation 
can be assumed to be valid when the quadratic Zeeman term is much larger than the other energy scales, $q\gg \epsilon_k,|g_2| n$.  
The eigenvalues of the resulting time-independent matrix can be calculated analytically, but  
they will not be presented here as they have a very complicated form. 
The eigenvalues show that a sodium condensate is always stable against long wavelength 
perturbations, which is in agreement with the results of the previous subsection 
if $q \gg g_2n$. Rubidium condensate has unstable states, and  
the largest region of instability in the $(\epsilon_k,f^2)$ plane is obtained by choosing  $\beta=\frac{\pi}{2}$ and $\gamma=0$. The kinetic energy of the unstable plane waves 
is bound by the condition $\epsilon_k\leq 2|g_2|n$. 
The eigenvectors of the Bogoliubov matrix and the  corresponding 
perturbations $\delta\psi$ were obtained 
numerically. There exists always two density modes $\delta\psi^{1,2}$, which can  
 approximately be written as $\delta\psi^{1,2}\approx C \psi$, where $C$ is a 
time- and position -dependent function. The density modes are stable. 
 The remaining four modes $\delta\psi^{3,4,5,6}$ are approximately orthogonal to $\psi$, 
but it is not as easy to characterize these  
modes as in the case where the spin and magnetic field are parallel 
($\beta=0$). In general, all these modes affect both magnetization and spin direction. 
However, when $\beta =\pi/2$, these modes can be classified into magnetization and 
spin modes. The magnetization mode is of the form (\ref{psi34}) with $f_z=0$. This mode 
changes, in addition to the magnetization, also the spin component in the $xy$ plane.  
The spin mode does not change the direction of the spin but only its amplitude $f$. 
In Fig.\ \ref{fig:limit} we plot the positive imaginary part $\omega_\textrm{i}$ 
of the eigenvalues of these modes. 
\section{Spin and magnetic field orthogonal}\label{sec:orthogonal}
 
\begin{figure}[h]
\begin{center}
\includegraphics[scale=.8]{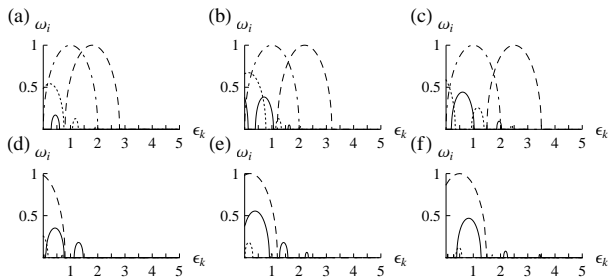} 
\end{center}
\caption{The unstable modes of (a)-(c) rubidium and (d)-(f) sodium for $\psi_\perp$ ($f=1$) and $\psi_\parallel$  ($f=0$). Here in (a) and (d) $q=0.8$, in (b) and (e) $q=1.2$, and in (c) and (f) $q=1.5$ in units of $|g_2|n$. The units of $\epsilon_k$ and $\omega_\textrm{i}$ are $|g_2|n$  and $|g_2|n/\hbar$, respectively. The dashed (dot-dashed) line gives the spin (magnetization) mode of  $\psi_\parallel$, while the solid (dotted) line indicates the spin (magnetization) mode of $\psi_\perp$. 
\label{fig:orthogonal}}
\end{figure}

In this section we compare the stability properties of 
states with $\f\parallel\mathbf{B}$ and $\f\perp\mathbf{B}$. 
 We argue that the energies of unstable plane waves for states with 
 $\f\nparallel \mathbf{B}$  are almost always smaller than the corresponding energies of the $\f\parallel\mathbf{B}$ case. 
This claim is based on energetic arguments. The kinetic energy $\epsilon_k$ 
of the domain structure can be assumed to increase as the energy  
of the initial state (with fixed magnetization) increases. 
The energy of the Zeeman term, $q(1-\rho_0)$, 
is maximized when $\rho_0=0$, which is the case if and only if the initial state is  $\psi_\parallel$. 
Furthermore, for a rubidium condensate also the interaction 
energy is maximized by $\psi_\parallel$ because then $g_2n f^2/2=-|g_2| n f_z^2/2$, which is the 
largest possible spin interaction energy for a homogeneous state with magnetization $f_z$. For sodium the situation is more complicated. 
For $q\gg g_2 n$ the magnetic-field energy dominates and $\psi_\parallel$ maximizes the energy. 
On the other hand, if $q < g_2 n$, the energy is maximized when $f\approx 1$. As in the case $\mathbf{B}\parallel \f$, states corresponding to the largest possible kinetic energy of the domain structure can be expected to be those with $f_z=0$. Therefore in the following we assume that magnetization vanishes.   
It is easy to show that  under this condition $\psi_\parallel$ (with $f_z=0$) is the state with  highest energy if $q\geq 2g_2n$. 
On the other hand, when $q=0$, the energy is maximized by 
\begin{align}
\label{psiperp}
\psi_\perp=
\frac{\sqrt{n}}{2}
\begin{pmatrix}
1 \\ \sqrt{2} \\ 1
\end{pmatrix},
\end{align}    
for which $f=1$ and which is unique up to a global phase and a rotation around the $z$ axis. 
We now compare the stability of this state to that of $\psi_\parallel$. 
 Numerically, it can be shown that for this state the operator $\hat{H}_B$ is periodic  
 and it is therefore possible to use Floquet analysis to study the stability. 
The Floquet theorem (see, e.g., \cite{Chicone})  states that if $\hat{H}_B$ is periodic,  
the time evolution operator $\U_{B}$ determined by equation (\ref{HBG}) 
can be written as 
\begin{align}
\U_{B}(t)=\hat{M}(t) e^{-i t \hat{K}},
\end{align}
where $\hat{M}$ is a periodic matrix with period $T$ and $\hat{M}(0)=\textrm{I}$ 
and $\hat{K}$ is some time-independent matrix. 
At times $t=nT$, where $n$ is an integer, we get $\U_{B}(nT)=e^{-i n T \hat{K}}$. The eigenvalues of $\hat{K}$ determine the stability of the system. If $\U_{B}(T)$ were unitary, all the eigenvalues 
of $\hat{K}$ would be real. In our case  $\U_{B}(T)$ does not have to be 
unitary and the eigenvalues of $\hat{K}$ can have a nonvanishing imaginary part. 
We say that the system is unstable if at least one of the eigenvalues of $\hat{K}$ has a positive imaginary part. We denote the imaginary part of an eigenvalue $\omega$ of $\hat{K}$ by $\omega_{\textrm{i}}$ and calculate it from 
\begin{align}
\omega_\textrm{i}=\frac{\textrm{Im} [i\ln\lambda]}{T}, 
\end{align}
 where $\lambda$ is an eigenvalue of $\U_{B}(T)$. 
 We calculated the eigenvalues and eigenvectors  
 numerically for various values of $q$. The oscillation period $T$ can be obtained from the equations given in \cite{Mur-Petit09}. 
 The unstable perturbations corresponding to the eigenvectors of $\hat{K}$ are similar 
 to the ones obtained in the previous section in the $\beta=\pi/2$ case. 
 Hence the magnetization mode changes both magnetization and 
 the direction and length of the spin vector $\f$ and the spin mode affects only the length 
 of the spin vector. In Fig.\ \ref{fig:orthogonal} we plot the unstable modes for some values of $q$. For comparison, also the unstable modes of the $\psi_\parallel$ states are shown. 
We find that for rubidium the maximal kinetic energy of the unstable perturbations of $\psi_\parallel$ is always higher than that of $\psi_\perp$. For sodium the same conclusion holds 
when $q\gtrsim 1.5 g_2n$. If $q\lesssim 1.5 g_2 n$ the maximal value of $\epsilon_k$ of
 can be slightly larger for $\psi_\perp$, as can be seen from Figs. \ref{fig:orthogonal}(d)-\ref{fig:orthogonal}(f). 
On the other hand, the growth rate of these instabilities is much smaller than the growth rate of the instabilities of $\psi_\parallel$. 
We therefore conclude that a lower bound for the wavelengths of unstable 
perturbations is essentially given by the equation $\epsilon_k=|g_2| n-g_2 n+q$,  
which is the corresponding bound for the states of the form $\psi_\parallel$.  
Consequently, we conjecture that for condensate sizes smaller than the wavelength  
corresponding to $\epsilon_k=|g_2| n-g_2 n+q$ both rubidium and sodium condensates are essentially 
stable regardless of the initial state. This wavelength is determined by
\begin{align}
\label{eq:lambda}
\lambda = \frac{2\pi\hbar}{\sqrt{2m(|g_2| n-g_2 n+q)}},
\end{align}
and wavelengths smaller than this are shown by the shaded region in Fig.\ \ref{fig:wavelengths}. One should note that  Eq. (\ref{eq:lambda}) gives only a sufficient condition for stability,  
 it does not allow us to conclude that a condensate is unstable if it is larger than this size. Depending on the initial state, the condensate may be stable even if it is larger than the size determined by (\ref{eq:lambda}).

In addition to giving a bound for stable condensate size, this result makes it possible to derive constraints for the validity of the single-mode approximation (SMA). The SMA states that spatial degrees of freedom decouple from spin dynamics when the condensate 
is smaller than the spin healing length 
\begin{align}
\label{eq:SMA}
\xi_s\equiv \frac{2\pi\hbar}{\sqrt{2m|g_2|n}}. 
\end{align} 
This condition is obtained by requiring that the spin-interaction energy is insufficient to create spatial spin structures and its 
 validity has been confirmed experimentally: 
For a ${}^{23}$Na condensate with Thomas-Fermi radius smaller than $\xi_s$, the 
SMA was found to provide a very good description of the system \cite{Black07}. 
However, the validity of SMA is also constrained by the results of the stability analysis discussed in this paper.  
If we assume that SMA holds initially, then the stability analysis shows that an additional 
requirement for the validity of SMA is that the condensate is smaller than the wavelength 
given by Eq. (\ref{eq:lambda}). In particular, at a high magnetic field ($q\gg |g_2|n$)  condition (\ref{eq:lambda}) gives a stricter bound for the condensate size than Eq. (\ref{eq:SMA}).  
We remark that an equation resembling Eq. (\ref{eq:lambda}) can be obtained also by equating 
the maximal energy in a magnetic field, $g_2n/2+q$, with the kinetic energy $\epsilon_k$. 
This is an extension of the argumentation used in obtaining Eq. (\ref{eq:SMA}) to the case where 
magnetic field is nonzero. The difference between these approaches is that Eq (\ref{eq:lambda}) is obtained from rigorous stability analysis, while the healing length argumentation is an order of magnitude estimate for the energy scales of the dynamics. 

\section{Bosons on a ring} \label{sec:ring}

As a specific realization of the instabilities discussed in this paper 
we study a gas of bosonic 
atoms in a toroidal trap. We consider a doughnut-shaped condensate with 
$N$ atoms, thickness $2\rho_{\perp}$ ($2\rho_z$) in the radial (axial) direction, 
and mean radius $R$, and assume that the 
trap is well approximated by a harmonic oscillator potential in the 
radial and axial directions, with trapping frequencies
 $\omega_{\perp}$ and $\omega_{z}$, respectively.  
Provided that both the spin healing length $\xi_\textrm{s}$ and the wavelength given by Eq. (\ref{eq:lambda}) are larger than $\rho_\perp$ and $\rho_z$, 
the SMA applies in radial and axial directions. This makes it possible to integrate out the 
dynamics in these directions.  
If, in addition, $R$ is large enough compared to $\rho_\perp$ and $\rho_z$, the condensate can be described as a homogeneous one-dimensional 
system of length $2\pi R$ with periodic boundaries. 
As a specific example, we discuss an optical trap of the type used in 
Ref.\ \cite{Ramanathan11}, created as a combination of a Laguerre-Gaussian 
beam and a laser sheet. The effective interaction energy is 
\begin{align}
g_2 n_{\textrm{eff}} &= \frac{N\hbar^2 (a_2-a_0)}{3m R \rho_{\perp} \rho_z}\frac{8}{3\pi},
\end{align}
where $n_{\textrm{eff}}$ comes from integrating the 
squared density in the Thomas-Fermi approximation in radial and axial directions. 
The Thomas-Fermi approximation can be assumed to be valid if $\hbar\omega_z,\hbar\omega_\perp\ll g_0 n_{\textrm{eff}}$. 
 
As we have seen, the parameters determining instabilities are 
$f$, the angles $\beta$ and $\gamma$, and the mode energy in units 
of the spin -interaction energy, $\epsilon_k/|g_2|n$. In the periodic 
geometry considered here, $k$ is quantized as $k=\kappa/R$, 
 where $\kappa$ is an integer. The corresponding  
 mode energy is $\epsilon_{\kappa}=\hbar^2\kappa^2/2mR^2$,  and 
 the allowed values for the ratio of the mode energy to the
  interaction energy are
\begin{align}
\label{allowed}
\frac{\epsilon_{\kappa}}{|g_2| n_{\textrm{eff}}} 
& = \frac{9\pi}{16}\frac{\rho_\perp \rho_z}{N R |a_2-a_0|}\kappa^2 \equiv 
e_1 \kappa^2,
\end{align}
where for convenience we introduced the dimensionless prefactor $e_1$; 
note that $e_1=\epsilon_1/|g_2|n_{\textrm{eff}}$. 
The characteristic time scale for the instabilities is seen 
from Eqs. (\ref{omega12})-(\ref{omega56}) and Fig.\ \ref{fig:parallel} 
to be given by $\hbar/|g_2|n_{\textrm{eff}}$ (note that the maximum magnitude of the spin and magnetization modes 
is independent of the magnetic-field parameter $q$).
With the chosen parameters, the time scale is about 130~ms for Rb and 10~ms for Na. 
We simulate the time development of the system starting from initial states of the form 
$\psi_\parallel$, which we argued to be the most unstable ones for given magnetization. 
We discuss first the time evolution of a rubidium condensate.  
\subsection{Rubidium} 
\begin{figure}
\includegraphics[width=0.45\textwidth]{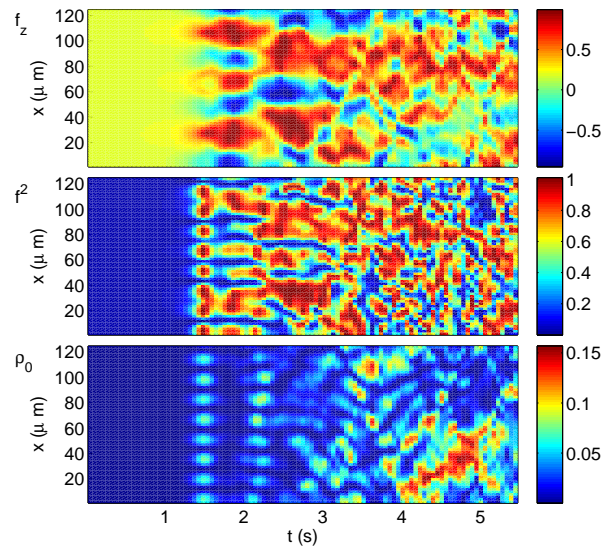}
\caption{(Color online) Time development of (top) the local spin projection $f_z$, (middle)  
squared spin $f^2$, and (bottom) zeroth component $\rho_0$ in a one-dimensional (1D) 
rubidium condensate with periodic boundary conditions. 
The system consists of 
$N=10^5$ Rb atoms in magnetic field $B=130$~mG ($q=|g_2|n_\textrm{eff}$), initially in state $\psi_\parallel$ [Eq. (\ref{psipara})] with $f=f_z=0.2$. 
The spatial dimension is along the vertical, and time is along the horizontal direction. 
}
\label{fig:sim1}
\end{figure}

Figure \ref{fig:sim1} displays the time development for $^{87}$Rb atoms 
in the initial state $\psi_\parallel$ with $f=0.2$ 
and in a magnetic field $B=130$~mG, corresponding to $q=|g_2|n_{\textrm{eff}}$, 
as in Fig.\ \ref{fig:parallel}(b). 
Note that since $\rho_0$ and $f^2$ depend on squared wave functions, the plots 
exhibit second harmonics, i.e., the number of peaks is twice the wave number 
$\kappa$ of the excitation. 
We see that the local spin amplitude $f$ and the population of the zero component 
$\rho_0$ develop instabilities with dominant wave number $\kappa=4$,  
corresponding to $\epsilon_4/|g_2|n_{\textrm{eff}}\approx 1.91$. This is close 
to the value $\epsilon_k/|g_2|n_{\textrm{eff}}=2$ which gives the fastest-growing spin mode; 
see Fig. \ref{fig:parallel}(b). Hence this mode is a spin mode. 
Another mode with wave number $\kappa=3$ affects both $f$ and 
the local magnetization $f_z$ but not the zeroth spin component, 
indicating that this is a magnetization mode. For this mode 
$\epsilon_4/|g_2|n_{\textrm{eff}}\approx 1.07$. As can be seen from Fig. \ref{fig:parallel}(b), this is close to the fastest-growing magnetization mode. 
We see that the linear analysis 
explains well the initial growth of the instabilities.   
At longer times, nonlinear processes take over. These will not  
be discussed in more detail here.  
The particle density, not plotted in Fig.\ \ref{fig:sim1}, stays constant to 
within a few percent; the instability only affects the spin. 
The time scale for buildup of an appreciable spin magnitude is slightly 
above 1 s, within which the modes have increased by about four 
orders of magnitude. This time is 
within attainable condensate lifetimes.

\begin{figure}
\includegraphics[width=0.45\textwidth]{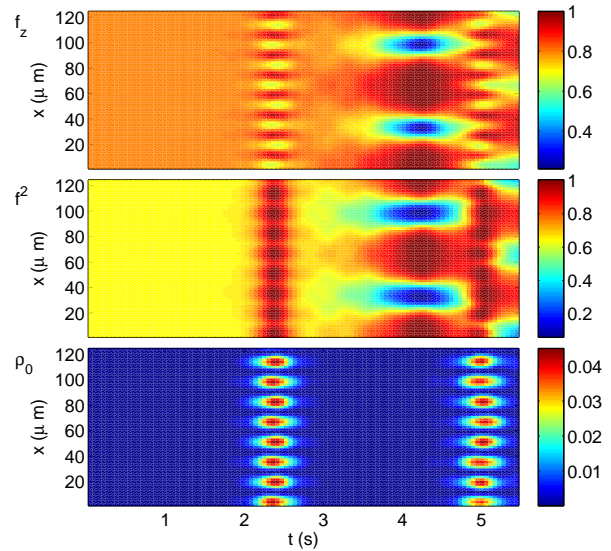}
\caption{(Color online) Time development of a 1D  rubidium condensate with periodic boundary conditions, 
as in Fig.\ \ref{fig:sim1}. Here the initial state has a magnetization 
$f_z=f=0.8$.}
\label{fig:sim2}
\end{figure}

If the initial state has a higher value of $f$, the wave number and the 
amplitude of the most unstable magnetization mode are decreased; 
see Fig.\ \ref{fig:parallel}~(b). An example for $f=f_z=0.8$ is given in 
Fig.\ \ref{fig:sim2}.
The spin mode still has wavenumber $\kappa=4$, which is consistent with the fact that the 
location of the fastest-growing spin mode does not depend on $f$.  
The wave number of the 
most unstable magnetization mode is reduced to $\kappa=2$. This gives 
 $\epsilon_2/|g_2|n_{\textrm{eff}}\approx 0.48$, while the fastest-growing 
 magnetization mode can be calculated from Eq. (\ref{omega34}) to be at $\epsilon_k/|g_2|n_{\textrm{eff}}\approx 0.36$.  Now it takes about 4 s for the instability to build up. 

\begin{figure}
\includegraphics[width=0.45\textwidth]{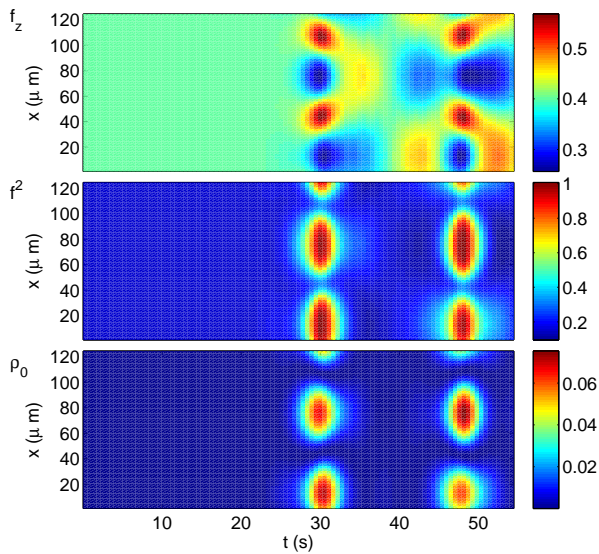}
\caption{(Color online) Time development of a 1D rubidium condensate with periodic boundary conditions, 
as in Fig.\ \ref{fig:sim1}. Here the system is made smaller so that the 
parameter $e_1=2$, and the external magnetic field is $B=10$~mG, 
corresponding to $q=|g_2|n_{\textrm{eff}}$. The initial state has a magnetization $f_z=f=0.4$.}
\label{fig:sim3}
\end{figure}

Assume next that the magnetic field vanishes and the trap 
parameters are tuned so that $e_1=2$. From Eq.\ (\ref{allowed}) we see that 
this can be done by, e.g., loosening the ring trap 
and decreasing the number of particles. 
Then the lowest modes, located at $\kappa=0$ and 
$\kappa=1$, give  $\epsilon_0/|g_2|n_{\textrm{eff}}=0$ and  $\epsilon_1/|g_2|n_{\textrm{eff}}=2$. 
Comparison with Fig. \ref{fig:orthogonal} shows that now all states $\psi_\parallel$ are stable. 
This is also what we see in the simulations (not shown here). 
However, by increasing the magnetic field we may once 
again make the system unstable. In Fig.\ \ref{fig:sim3} we report on a 
simulation where $e_1=2$ and $q=|g_2|n_{\textrm{eff}}$, corresponding to 
$B=10$mG if the radius $R$ is left unchanged. In the initial state  
$f_z=f=0.4$.
In such a magnetic field, we expect the spin mode to be the only unstable mode, 
with wave number $\kappa=1$. The increase in $\rho_0$  caused by the spin mode 
is clearly visible in Fig.\ \ref{fig:sim3}. 
An oscillation with wave number $\kappa=2$ is seen to develop in the magnetization 
simultaneously; this is not predicted by the linear analysis  
since $\kappa=2$ lies outside the unstable region in this 
case. However, a closer look at the Fourier transform of the 
spin components shows that 
this is not due to a linear instability but is a nonlinear effect. 
\begin{figure}
\includegraphics[width=0.45\textwidth]{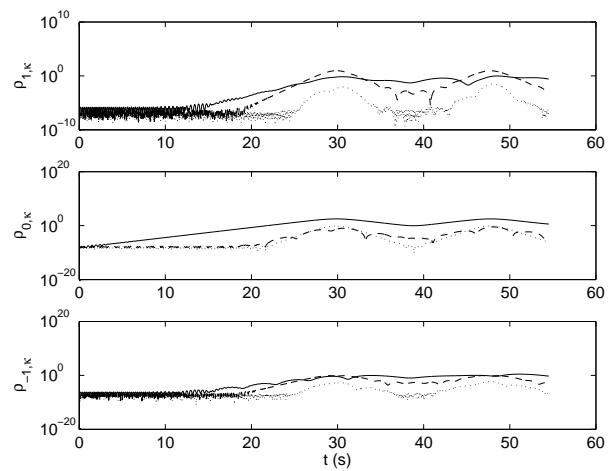}
\caption{Populations of the lowest plane-wave components of the system 
in Fig.\ \ref{fig:sim3}. (top) Populations $\rho_{1,\kappa}$ of 
spin component $m_F=1$, (middle) spin component $m_F=0$, and (bottom) 
spin component $m_F =-1$. Solid lines show $\kappa=1$, dashed lines show  
$\kappa=2$, and dotted lines show $\kappa=3$.
\label{fig:fftsim}}
\end{figure}
In Fig. \ref{fig:fftsim}, we see an exponential rise of the 
population $\rho_{0,1}$, i.e., the $\kappa=1$ plane wave component 
of the $m_F=0$ spin component. Populations in the $m_F=\pm 1$ components, 
both in $\kappa=1$ and $\kappa=2$, are excited as secondary instabilities.  

\subsection{Sodium}
\begin{figure}
\includegraphics[width=0.45\textwidth]{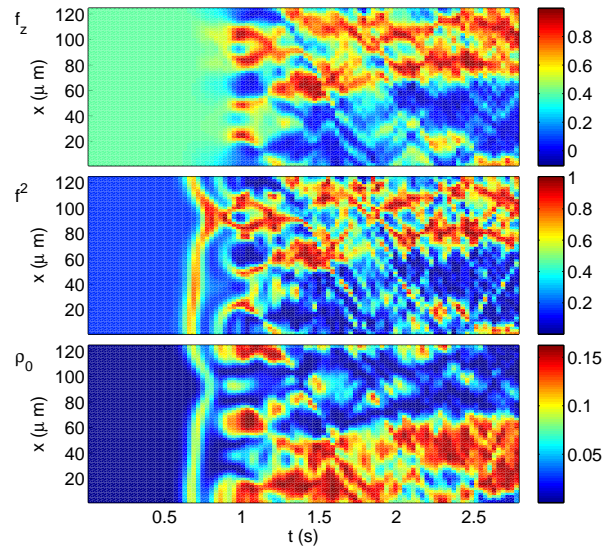}
\caption{(Color online) Time development of a 1D  sodium condensate with periodic boundary conditions, 
as in Fig.\ \ref{fig:sim1}. Here we simulate ${}^{23}$Na atoms in an 
initial state with magnetization $f=0.4$, and the magnetic field is 
 $B=95$~mG ($q=g_2 n_{\textrm{eff}}/2$).}
\label{fig:sim4}
\end{figure}

\begin{figure}
\includegraphics[width=0.45\textwidth]{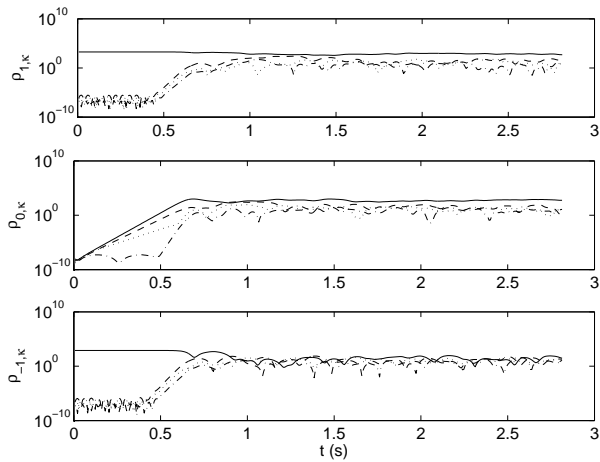}
\caption{Populations of the lowest plane-wave components of the system 
in Fig.\ \ref{fig:sim4}. Panels are as in Fig.\ \ref{fig:fftsim}.
Solid lines show $\kappa=0$, dashed lines show  
$\kappa=2$, dotted lines show $\kappa=3$, and dash-dotted lines show $\kappa=4$.
\label{fig:fftsim2}}
\end{figure}

We now consider a system of ${}^{23}$Na atoms in a toroidal trap 
with the same Thomas-Fermi length parameters as above. 
For these parameters, $e_1=0.033$.
The system is stable in zero field, as seen in Fig.\ 1(d). If, on the 
other hand, $q=g_2n_{\textrm{eff}}/2$ ($B=95$~mG) [cf.\ Fig.\ 1(e), 
where $q=g_2n_{\textrm{eff}}$], the main instability develops 
at $\kappa=0$. 
The result of the simulation is shown in Fig.\ \ref{fig:sim4}. 
For this simulation we chose an initial state with $f=0.4$, 
 which allows instabilities with $\kappa=0,1,2,3$.   
Indeed, instabilities now develop with wave numbers from $\kappa=0$ 
up to 3. This is more clearly 
seen in the plot of the Fourier components in Fig.\ \ref{fig:fftsim2} 
 (where the $\kappa=1$ component, whose time dependence is similar to 
that of the $\kappa=2$ and $\kappa=3$ components, is left out in order not to clutter 
the figure).
It is seen that the most unstable mode has wavenumber $\kappa=0$ and 
corresponds to uniformly populating the $m=0$ spin component.

The results reported in this section indicate that the dynamical 
instabilities studied in Secs.\ \ref{sec:analytical} and \ref{sec:orthogonal} 
can be readily studied in 
existing traps and that the wave number of the unstable modes can 
be controlled by managing the system size and magnetic field. 
Systems small enough to be stable seem to be within reach. 
Time scales are also clearly tunable.

\section{Conclusions}\label{sec:conclusions}
We have studied the stability of spin-1 Bose-Einstein 
condensates, concentrating on the nonstationary states of 
rubidium and sodium condensates. The analysis was performed in a frame of reference where the state under investigation is stationary. The stability analysis was done 
using the Bogoliubov approach, that is, expanding the time evolution equations of the system 
to first order with respect to a small perturbation in the stationary state wave function. 
The resulting time evolution equations for the perturbations were solved analytically and numerically, assuming that the unperturbed system is spatially homogeneous. 
In particular, the effect of an external homogeneous magnetic field was examined. 
We found that the eigenmodes and eigenvectors of the perturbations can be determined analytically if the spin and magnetic field are parallel, regardless of the strength 
of the magnetic field. These eigenmodes show that a 
$^{87}$Rb condensate has long-wavelength instabilities which are independent of the 
strength of the magnetic field. These do not exist in a $^{23}$Na condensate. 
Additionally, instabilities whose wavelengths depend on the strength of the magnetic field 
are possible in both systems. For rubidium these exist already at zero field, while 
for sodium nonzero magnetic field is required for the instability to appear.

The stability of long wavelength perturbations was solved analytically also in the 
case where the magnetic -field energy is much larger than the spin 
interaction energy and the kinetic energy of the plane wave perturbations.  
The wavelengths of the unstable long wavelength perturbations are bounded by 
the condition $\epsilon_k\leq 2|g_2|n$ regardless of the initial state.  

It was also argued that states with spin parallel to the magnetic field are the ones 
whose instabilities have the highest energy. This claim was based on energetic arguments and 
a numerical study of the stability of a state that is orthogonal to the magnetic field. 
The results allow us to derive an analytical formula giving a sufficient condition for 
the size of a stable condensate at a given magnetic field. Condensates smaller than the   
size given by Eq. (\ref{eq:lambda}) are guaranteed to be stable. However, all condensates 
larger than this are not necessarily  unstable; if prepared in a suitable state, the 
system may be stable even if it is larger than this size.  Equation (\ref{eq:lambda})    
gives also a criterium for the validity of the single-mode approximation.  
At a high magnetic field this condition gives a stricter bound for the condensate size than the 
standard condition, given by the spin healing length. 
  
Finally, the stability properties predicted by the linear Bogoliubov theory were studied  
by solving the Gross-Pitaevskii equations numerically in a 1D circular geometry. 
It was shown that by controlling the number of particles, trapping frequencies, and 
strength of the magnetic field it is possible to control the stability properties of the condensate.

\begin{acknowledgments}
The authors thank Luis Santos for helpful discussions. 
M.J. and E.L. acknowledge financial support from the Swedish Research Council. 
\end{acknowledgments}

\begin{appendix}
\section{}
\label{appendixA}
Here we examine the time evolution of spin states 
by looking at the time evolution equations of the system. 
An arbitrary spin state can be written as 
\begin{align}
\label{generalpsi}
\psi=\sqrt{n}
\begin{pmatrix}
e^{i\theta_1}\sqrt{\frac{1}{2}(1-\rho_0+f_z)}\\
e^{i\theta_0}\sqrt{\rho_0}\\
e^{i\theta_{-1}}\sqrt{\frac{1}{2}(1-\rho_0-f_z)}
\end{pmatrix}. 
\end{align}
Writing $\psi$ in this way and using Eq. (\ref{time-evolution}) 
give the time evolution equations
\begin{align*}
\hbar\frac{\partial \rho_1}{\partial t} &=\hbar\frac{\partial \rho_{-1}}{\partial t}
=-\frac{\hbar}{2}\frac{\partial \rho_0}{\partial t}
=g_2 n\rho_0\sqrt{(1-\rho_0)^2-f_z^2}\sin\Theta,\\
\hbar\frac{\partial \theta_{\pm 1}}{\partial t} &=-g_2 n
\left(\rho_0\sqrt{\frac{1-\rho_0\mp f_z}{1-\rho_0\pm f_z}}\cos\Theta+\rho_0\pm f_z\right)-q\pm p,\\
\hbar\frac{\partial \theta_{0}}{\partial t} &=-g_2 n\left(\sqrt{(1-\rho_0)^2-f_z^2}\cos\Theta+1-\rho_0\right),\\
\hbar\frac{\partial }{\partial t}\Theta
&=2g_2 n\left(\frac{(1-\rho_0)(2\rho_0-1)+f_z^2}{\sqrt{(1-\rho_0)^2-f_z^2}}\cos\Theta+2\rho_0-1\right)+2q,\\
\Theta &=2\theta_0-\theta_1-\theta_{-1}.  
\end{align*}
In deriving these equations we have neglected the term proportional to the identity operator as it changes only the global phase.   
Clearly, if $\rho_0=0$ in the initial state, the populations will remain constant during the 
subsequent time evolution. This means that only the phases of the state $\psi_\parallel$, given in Eq. (\ref{psipara}),  
can evolve in time. Another special case is obtained when $f_z=0$, which corresponds to the spin vector lying in the $xy$ plane. 
In this case $\theta_1(t)=\theta_{-1}(t)$ [assuming that $\theta_1(0)=\theta_{-1}(0)$]. 
Because the time evolution of $\Theta$ and $\rho_0$ is periodic (modulo $2\pi$) with the same period, also the time evolution of the state vector is periodic, up to a global phase. 
This can be seen by redefining the phases as $\theta_k'(t)=\theta_k(t)-\theta_1(t)$, which gives $\theta_1'(t)=\theta_{-1}'(t)=0,\theta_0'(t)=\Theta(t)/2$. Although the state vector is periodic in time,  numerical calculations show that in general the Bogoliubov matrix $\hat{H}_B$ 
is not periodic. An exception is given by the state $\psi_\perp$. For this state 
$\hat{H}_B$ is periodic and the stability of $\psi_\perp$ can be analyzed using Floquet theory.  

It is possible to obtain an approximate propagator for state (\ref{generalpsi}) under the assumption 
that $q\gg |g_2|n$. Then $\Theta(t) \approx \Theta(0)+ 2qt/\hbar$, which leads to rapidly oscillating $\sin\Theta$ and $\cos\Theta$ and we can average over one oscillation period, obtaining $\sin\Theta  \approx\cos\Theta \approx 0$. 
This gives $\dot{\rho_k}=0$, and we get the propagator 
\begin{align}
\hat{U}_\psi= 
e^{-i t g_2 n(1-\rho_0)/\hbar}e^{-i t[(g_2 n f_z-p) \hat{F}_z+(g_2 n(2\rho_0-1)+q)\hat{F}_z^2]/\hbar}.
\end{align}

\section{Eigenvectors}
\label{appendixB}
In the case where the magnetic field and spin are parallel the eigenvectors of $\hat{H}_B$  
can be calculated analytically and are given, up to a normalization, by
\begin{align}
\label{x1234}
\mathbf{x}_j &=(\alpha_j\,(\epsilon_k+\hbar\omega_j),0,\epsilon_k+\hbar\omega_j,
\alpha_j\,(\epsilon_k-\hbar\omega_j),0,\epsilon_k-\hbar\omega_j),\\
\nonumber
\mathbf{x}_j &= 
(0,g_2 n\sqrt{1-f^2}e^{iqt/\hbar},0,0,0,0)\\
\label{x56}
&+(0,0,0,0,(-\epsilon_k-g_2 n+q+\hbar\omega_j)e^{-iqt/\hbar},0).
\end{align} 
Here in the first equation $j=1,2,3,4$ and in the second one $j=5,6$, and 
\begin{align}
\alpha_j \equiv \frac{f(g_0+g_2)+s_j \sqrt{(g_0-g_2)^2+4g_0 g_2 f^2}}{(g_0-g_2)\sqrt{1-f^2}},
\end{align}
where we have defined $s_1=s_2=-s_3=-s_4=1$. 
The corresponding perturbations become

\begin{align}
\label{pert1}
\delta\psi^{j}&= \sum_{\mathbf{k}}  C_j F
\begin{pmatrix}
\alpha_j\\
0\\
1
\end{pmatrix},
\end{align}
where 
\begin{align}
\label{pert2}
F = 
\left\{
\begin{array}{ll}
\hbar\omega_j\cos(\mathbf{k}\cdot\mathbf{r}+\omega_j t)
+i\epsilon_k \sin(\mathbf{k}\cdot\mathbf{r}+\omega_j t), &\omega_j\in\mathbb{R} \\
(\mp \hbar |\omega_j|+i\epsilon_k)e^{\mp |\omega_j| t} 
 \sin(\mathbf{k}\cdot \mathbf{r}),  & \omega_j=\pm i |\omega_j|,
\end{array}\right.
\end{align}
and $C_j$ is an arbitrary nonzero complex number and $ j=1,2,3,4$. For $j=5,6$ we get  
\begin{align}
\label{pert3}
\nonumber
\delta\psi^{j} &= \sum_{\mathbf{k}} C_j e^{iqt/\hbar}\Big[g_2 n\sqrt{1-f^2}
e^{i(\mathbf{k}\cdot\mathbf{r}+\omega_j t)}\\
&-(-\epsilon_k-g_2 n+q+\hbar\omega_j)e^{-i(\mathbf{k}\cdot\mathbf{r}+\omega_j t)} \Big]
\begin{pmatrix}
0\\
1\\
0
\end{pmatrix}.
\end{align}
In order to derive an approximate expression 
for $\delta\psi^j$, we expand $\alpha_j$ in Taylor series with respect to $g_2/g_0$. We get 
\begin{align}
\alpha_j=\frac{f+s_j}{\sqrt{1-f^2}}\left[1+\mathcal{O}(g_2/g_0)\right].
\end{align} 
For rubidium and sodium $|g_2|/g_0 \ll 1$, which allows us to 
include only the zeroth order term in the above equation. This gives 
\begin{align}
\nonumber
\delta\psi^{j} &= \frac{s_j C_j F}{\sqrt{(1-s_j f)/2}} 
\begin{pmatrix}
\sqrt{(1+s_j f)/2}\\
0\\
s_j\sqrt{(1-s_j f)/2}
\end{pmatrix}.
\end{align}
Here $j=1,2,3,4$.   
\end{appendix}


\begin{thebibliography}{2}%
\makeatletter
\providecommand \@ifxundefined [1]{%
 \@ifx{#1\undefined}
}%
\providecommand \@ifnum [1]{%
 \ifnum #1\expandafter \@firstoftwo
 \else \expandafter \@secondoftwo
 \fi
}%
\providecommand \@ifx [1]{%
 \ifx #1\expandafter \@firstoftwo
 \else \expandafter \@secondoftwo
 \fi
}%
\providecommand \natexlab [1]{#1}%
\providecommand \enquote  [1]{``#1''}%
\providecommand \bibnamefont  [1]{#1}%
\providecommand \bibfnamefont [1]{#1}%
\providecommand \citenamefont [1]{#1}%
\providecommand \href@noop [0]{\@secondoftwo}%
\providecommand \href [0]{\begingroup \@sanitize@url \@href}%
\providecommand \@href[1]{\@@startlink{#1}\@@href}%
\providecommand \@@href[1]{\endgroup#1\@@endlink}%
\providecommand \@sanitize@url [0]{\catcode `\\12\catcode `\$12\catcode
  `\&12\catcode `\#12\catcode `\^12\catcode `\_12\catcode `\%12\relax}%
\providecommand \@@startlink[1]{}%
\providecommand \@@endlink[0]{}%
\providecommand \url  [0]{\begingroup\@sanitize@url \@url }%
\providecommand \@url [1]{\endgroup\@href {#1}{\urlprefix }}%
\providecommand \urlprefix  [0]{URL }%
\providecommand \Eprint [0]{\href }%
\providecommand \doibase [0]{http://dx.doi.org/}%
\providecommand \selectlanguage [0]{\@gobble}%
\providecommand \bibinfo  [0]{\@secondoftwo}%
\providecommand \bibfield  [0]{\@secondoftwo}%
\providecommand \translation [1]{[#1]}%
\providecommand \BibitemOpen [0]{}%
\providecommand \bibitemStop [0]{}%
\providecommand \bibitemNoStop [0]{.\EOS\space}%
\providecommand \EOS [0]{\spacefactor3000\relax}%
\providecommand \BibitemShut  [1]{\csname bibitem#1\endcsname}%
\let\auto@bib@innerbib\@empty
\bibitem [{Note1()}]{Note1}%
  \BibitemOpen
  \bibinfo {note} {This can be proven as follows. If $\protect \mathaccentV
  {hat}05E{V}$ is an element of the (now unitary) symmetry group of the energy,
  then $\protect \mathaccentV {hat}05E{U}_{\protect \mathaccentV
  {hat}05E{V}\psi }=\protect \mathaccentV {hat}05E{V}\protect \mathaccentV
  {hat}05E{U}_\psi \protect \mathaccentV {hat}05E{V}^{\protect \dag }$. By
  replacing $\psi (0)\rightarrow \protect \mathaccentV {hat}05E{V}\psi (0)$ and
  $\protect \mathaccentV {hat}05E{U}_\psi \rightarrow \protect \mathaccentV
  {hat}05E{V}\protect \mathaccentV {hat}05E{U}_\psi \protect \mathaccentV
  {hat}05E{V}^{\protect \dag }$ in Eqs. (\ref {X}) and (\ref {Y}) we find that
  $\protect \mathaccentV {hat}05E{X}\rightarrow \protect \mathaccentV
  {hat}05E{V} \protect \mathaccentV {hat}05E{X}\protect \mathaccentV
  {hat}05E{V}^\protect \dag $ and $\protect \mathaccentV {hat}05E{Y}\rightarrow
  \protect \mathaccentV {hat}05E{V} \protect \mathaccentV {hat}05E{Y}\protect
  \mathaccentV {hat}05E{V}^T$. Consequently, $\protect \mathaccentV
  {hat}05E{H}_B\rightarrow \protect \mathaccentV {hat}05E{W}\protect
  \mathaccentV {hat}05E{H}_B\protect \mathaccentV {hat}05E{W}^\protect \dag $,
  where $\protect \mathaccentV {hat}05E{W}$ is a block diagonal matrix
  $\protect \mathaccentV {hat}05E{W}=\protect \textrm {diag}(\protect
  \mathaccentV {hat}05E{V}\protect \tmspace +\thinmuskip {.1667em}\protect
  \tmspace +\thinmuskip {.1667em} \protect \mathaccentV {hat}05E{V}^*)$.
  Because $\protect \mathaccentV {hat}05E{H}_B$ and $\protect \mathaccentV
  {hat}05E{W}\protect \mathaccentV {hat}05E{H}_B\protect \mathaccentV
  {hat}05E{W}^\protect \dag $ have the same eigenvalues, they also have
  identical stability properties.}\BibitemShut {Stop}%
\bibitem [{Note2()}]{Note2}%
  \BibitemOpen
  \bibinfo {note} {In a sodium condensate, instead of creating domains, it may
  be preferable to increase $\rho _0$ (and thus also $f$) at a low magnetic
  field. This corresponds to a spin mode with $k=0$. This mode is not present
  in rubidium. This can be understood by noting that an increase in $\rho _0$
  leads to an increase in $f$. In a rubidium condensate this decreases the spin
  interaction energy, which, together with decreasing magnetic field energy,
  leads to an energy surplus that is transferred into kinetic energy. In a
  sodium condensate increasing $f$ increases the spin interaction energy. In
  some cases this matches exactly the energy released from the quadratic Zeeman
  term, producing a mode with $k=0$.}\BibitemShut {Stop}%
\end{thebibliography}%


\begin{thebibliography}{19}

\bibitem{Ho98} T.-L. Ho, Phys. Rev. Lett. {\bf 81}, 742 (1998)

\bibitem{Ohmi98} T. Ohmi and K. Machida , J. Phys. Soc. Jap. {\bf 67}, 1822 (1998).

\bibitem{Ueda00} M. Ueda, Phys. Rev. A {\bf 63}, 013601 (2000).

\bibitem{Ueda02} M. Ueda and M. Koashi, Phys. Rev. A {\bf 65}, 063602 (2002).

\bibitem{Robins01} N. P. Robins, W. Zhang, E. A. Ostrovskaya, and Y. S. Kivshar, Phys. Rev. A {\bf 64}, 021601(R) (2001). 

\bibitem{Murata07} K. Murata, H. Saito, and M. Ueda, Phys. Rev. A {\bf 75}, 013607 (2007).


\bibitem{Stenger98} J. Stenger, S. Inouye, D. M. Stamper-Kurn, H.-J. Miesner, A. P. Chikkatur, and W. Ketterle, Nature (London) {\bf 396}, 345 (1998). 

\bibitem{Chang04} M.-S. Chang, C. D. Hamley, M. D. Barrett, J. A. Sauer, K. M. Fortier,  W. Zhang, L. You, and M. S. Chapman, Phys. Rev. Lett. {\bf 92}, 140403 (2004).

\bibitem{Chang05} M.-S. Chang, Q. S. Qin, W. X. Zhang, L. You, and M. S. Chapman, Nat. Phys. {\bf 1}, 111 (2005).

\bibitem{Kronjager05} J. Kronj\"ager, C. Becker, M. Brinkmann, R. Walser, P. Navez, K. Bongs, and K. Sengstock, Phys. Rev. A {\bf 72}, 063619 (2005).

\bibitem{Black07} A. T. Black, E. Gomez, L. D. Turner, S. Jung, and P. D. Lett, Phys. Rev. Lett. {\bf 99}, 070403 (2007).

\bibitem{Kronjager10} J. Kronj\"ager, C. Becker, P. Soltan-Panahi, K. Bongs, and K. Sengstock, Phys. Rev. Lett. {\bf 105}, 090402 (2010).

\bibitem{Sadler06} L. E. Sadler, J. M. Higbie, S. R. Leslie, M. Vengalattore, and D. M. Stamper-Kurn, Nature (London) {\bf 443}, 312 (2006). 

\bibitem{Leslie09} S. R. Leslie, J. Guzman, M. Vengalattore, J. D. Sau, M. L. Cohen, and 
D. M. Stamper-Kurn, Phys. Rev. A {\bf 79}, 043631 (2009). 

\bibitem{Guzman11} J. Guzman, G.-B. Jo, A. N. Wenz, K. W. Murch, C. K. Thomas, and D. M. Stamper-Kurn, arXiv:1107.2672. 

\bibitem{Matuszewski08} M. Matuszewski, T. J. Alexander, and Y. S. Kivshar, Phys. Rev. A {\bf 78}, 023632 (2008). 

\bibitem{Matuszewski09} M. Matuszewski, T. J. Alexander, and Y. S. Kivshar, Phys. Rev. A {\bf 80}, 023602 (2009). 

\bibitem{Matuszewski10} M. Matuszewski, Phys. Rev. Lett. {\bf 105}, 020405 (2010). 

\bibitem{Zhang05} W. Zhang, D. L. Zhou, M.-S. Chang, M. S. Chapman and L. You, Phys. Rev. Lett. {\bf 95}, 180403 (2005). 

\bibitem{vanKempen02} E. G. M. van Kempen, S. J. J. M. F. Kokkelmans, D. J. Heinzen, and B. J. Verhaar, Phys. Rev. Lett. {\bf 88}, 093201 (2002).

\bibitem{Crubellier99} A. Crubellier, O. Dulieu, F. Masnou-Seeuws, M. Elbs, 
H. Kn\"ockel, and E. Tiemann, Eur. Phys. J. D. {\bf 6}, 211 (1999).

\bibitem{Burke98}J. P. Burke, C. H. Greene, and J. L. Bohn, Phys. Rev. Lett. {\bf 81}, 3355 (1998).

\bibitem{Gerbier06} F. Gerbier, A. Widera, S. F\"olling, O. Mandel, and I. Bloch, Phys. Rev. A {\bf 73}, 041602(R) (2006). 

\bibitem{Chicone} C. Chicone, {\it Ordinary Differential Equations with Applications} 
(Springer, New York, 1999).

\bibitem{Mur-Petit09} J. Mur-Petit, Phys. Rev. A {\bf 79}, 063603 (2009).

\bibitem{Ramanathan11} A. Ramanathan, K. C. Wright, S. R. Muniz, M. Zelan, W. T. Hill, III, 
C. J. Lobb, K. Helmerson, W. D. Phillips, and G. K. Campbell, Phys. Rev. Lett. {\bf 106}, 130401 (2011).   






\end{thebibliography}
\end{document}